\documentclass[twocolumn, superscriptaddress]{revtex4}
\usepackage{setspace}
\usepackage[german,american,english]{babel}
\usepackage{graphicx}
\usepackage{graphics}
\usepackage{dcolumn}
\usepackage{bm}
\usepackage{amssymb}
\usepackage{amsmath}
\usepackage{amsfonts}
\usepackage{epsfig}

\newcommand {\ket} [1] {| #1 \rangle}

\newcommand {\dbkt} [2] {\langle #1 | #2 \rangle}
\newcommand {\tbkt} [3] {\langle #1 | #2 | #3 \rangle}

\newcommand {\beq}{\begin{equation}}
\newcommand {\eeq}{\end{equation}}

\begin{document}

\title{Dephasing of Si spin qubits due to charge noise}
\author{Dimitrie Culcer}
\affiliation{Condensed Matter Theory Center, Department of Physics, University of Maryland, College Park MD20742-4111}
\author{Xuedong Hu}
\affiliation{Condensed Matter Theory Center, Department of Physics, University of Maryland, College Park MD20742-4111}
\affiliation{Joint Quantum Institute, Department of Physics, University of Maryland, College Park MD20742-4111}
\affiliation{Department of Physics, University at Buffalo, SUNY, Buffalo, NY 14260-1500}
\author{S.~Das Sarma}
\affiliation{Condensed Matter Theory Center, Department of Physics, University of Maryland, College Park MD20742-4111}
\affiliation{Joint Quantum Institute, Department of Physics, University of Maryland, College Park MD20742-4111}

\begin{abstract}
Spin qubits in Silicon quantum dots can have long coherence times, yet their manipulation relies on the exchange interaction, through which charge noise can induce decoherence. Charge traps near the interface of a Si heterostructure lead to fluctuations in the quantum-dot confinement and barrier potentials, which cause gating errors and two-spin dephasing. We quantify these effects in Si double quantum dots using a realistic model of noise. Specifically, we consider both random telegraph noise from a few traps (good for dots grown on submicron wafers) and $1/f$ noise from many traps (good for larger wafers appropriate for quantum dot arrays). We give estimates of gate errors for single-spin qubit architectures and dephasing in singlet-triplet qubits.
\end{abstract}
\maketitle

Solid-state spin-based qubits \cite{LDV} are believed to be promising for scalable quantum computation (QC).  Spins are weakly coupled to the environment leading to long spin coherence times, while semiconductor-based structures, such as quantum dots (QDs), can in principle be straightforwardly scaled up.  Among semiconducting host materials, Si stands out with small spin-orbit interaction, no piezoelectric electron-phonon coupling, and weak (which could
be further reduced through isotopic purification) hyperfine coupling, so that single electron spins have excellent quantum coherence properties, with a coherence time for donor electron spins in Si:P of up to 300 ms \cite{Lyon_Private}.  This coherence time is the longest among solid state qubits, and considerably exceeds the electron spin coherence time of $\sim 1 \mu$s in GaAs QDs \cite{PettaSci, Koppens_PRL08}, which is plagued by hyperfine interaction with nuclear spins.  Moreover, the sophisticated Si fabrication technology is a potential asset in scaling up Si-based QC architectures.  

At present, lateral quantum dots have been fabricated in Si structures such as Si/SiGe quantum wells \cite{Copper} and Si/SiO$_2$ heterojunctions \cite{Private}.  An important experimental and theoretical issue concerns the spin coherence properties in a single or double QD near a Si/SiO$_2$ interface, where the presence of defects such as $P_b$ centers is inevitable and the oxide is well known as a source of charge noise \cite{Helms_RPP94}.  Will the superior spin coherence properties in bulk Si hold up?  Even if charge fluctuations do not affect single spins significantly, when spins are coupled via exchange interaction, they can be adversely affected by charge noise because exchange is Coulombic in nature \cite{HuDS_06}.  Since the main perceived advantage of Si spin qubits is the expected long spin coherence time because of weak hyperfine coupling (and the potential for isotopic purification by removing Si-29 nuclei, thus further reducing the coupling to nuclear spins), it is crucial to consider the decoherence of Si spin qubits due to the electrostatic noise or fluctuations in the environment affecting the exchange coupling in the system, which is independent of nuclear spin considerations.

In this Letter we quantify the effect of charge noise on exchange-coupled spin qubits in Si double quantum dots (DQDs).  In particular we calculate two-qubit gate errors when single spins are used as qubits \cite{LDV}, and decoherence rates when two-spin singlet and triplet states serve as encoded qubit states \cite{PettaSci}.  Below we first give a brief discussion of charge noise near a Si/SiO$_2$ interface, then set up the DQD Hamiltonian.  We calculate the exchange coupling in the Heitler-London approximation and determine the modifications to the exchange due to variations in the barrier and level detuning.  Finally we discuss dephasing rates in singlet-triplet qubits.  

In a Si MOSFET structure dangling bonds ($P_b$ centers) near the interface act as charge traps, capturing electrons from a nearby source such as a 2-dimensional electron gas (2DEG) and re-emitting them.  Charge traps reside in the oxide layer up to a few nm from the interface, within tunneling distance from the 2DEG.  A trap is characterized by an activation energy $E_t$, which is the energy required to change the charging state of the trap.  For example, $P_b$ centers generally lie below the Fermi energy \cite{Campbell, Lenahan}, thus the traps are mostly occupied, so that $E_t - E_F$, where $E_F$ is the Fermi energy of the 2DEG, corresponds to the energy the trap needs to emit an electron to the Fermi surface of  the 2DEG.  The mean capture and emission times $\langle \tau_c \rangle$ and $\langle \tau_e \rangle$ of the traps satisfy $\langle \tau_c \rangle/ \langle \tau_e \rangle = e^{(E_t - E_F)/k_BT}$.  These times depend sensitively on temperature and gate voltage, and in a Si MOSFET at low temperatures they can range from $<$1 ms to $>$1 s \cite{Grenoble}.  The number of carriers in the trap typically fluctuates by one.  Charge fluctuations due to one trap are represented by a random function $\Delta V(t)$.  We take $\Delta V$ to have the same magnitude for all traps and neglect interactions between traps.  In a classical transistor, the alternate capture and emission of carriers by individual defect sites generates discrete switching events in the current through the device called random telegraph noise (RTN). The spectral density of RTN is 
\begin{equation}
S_V (\omega) = \frac{4 \Delta V^2}{(\langle \tau_e \rangle + \langle \tau_c \rangle)\big[\big(\frac{1}{\langle \tau_e \rangle} 
+ \frac{1}{\langle \tau_c \rangle}\big)^2 + \omega^2\big]}.
\end{equation}

In submicron devices there are typically only a few traps \cite{Wirth_IEEE05RTN}.  Yet scalable quantum computation requires the fabrication of large QD arrays.  In such devices, which can be $\gg 1 \mu$m in length scale, the ensemble of traps has a distribution of activation energies $E_t$, and thus have different capture and emission times.  Such a sum of many random telegraph signals yields a $1/f$ spectral density \cite{Dutta_RMP81, Kogan}, a fact well documented in MOSFETs \cite{Ralls_PRL84,Kogan} and other microelectronic devices \cite{Jung_APL04,Astafiev_PRL04}.  For example in Ref.~\cite{Fleetwood_IEEE02}, where $1/f$ noise was observed, the trap density $\sim$100 - 1000$\mu$m$^{-2}$, so 
that across a 100 nm section one expects a few fluctuators.  This is the case in small samples where discrete switchings are resolved \cite{Wirth_IEEE05RTN}.  For $1/f$ noise, as its name indicates, the spectral density takes the form $S_V(\omega) = \alpha k_BT/\omega$, where $\alpha$ represents the strength of this noise and a cutoff $\omega_0$ is introduced at the small-$\omega$ limit, often taken as the inverse of the measuring time.  In the present study we consider both RTN and $1/f$ noises.

To explore how charge noise affects exchange-coupled spin qubits in Si, we study a DQD in a 2DEG in a MOSFET structure grown along the $\hat{\bm z}$ direction.  The right and left dots are located at ${\bm R}_{R,L} = (\pm X_0, 0, 0)$ respectively.  The two-electron Hamiltonian is $H_0 = \big(\sum_{i=1,2}T^{(i)} + V_Q^{(i)} \big) + V_{ee}$, 
where $T$ is the kinetic energy and $V_Q$ the confinement potential, 
\begin{equation}
V_Q = \displaystyle V_x + \left( \frac{\hbar^2}{2m_ta^2} \right) \, 
\frac{y^2}{a^2} + \left(\frac{\hbar^2}{2m_zb^2}\right) \, \frac{z^2}{b^2},
\end{equation}
with $a$ and $b$ the Fock-Darwin radii for in- and out-of-plane confinement, and $m_t$ and $m_z$ the in- and out-of-plane Si effective masses.  $V_x$ is given below for symmetric and asymmetric dots.  The Coulomb potential between electrons at ${\bm r}_1$ and ${\bm r}_2$ is $V_{ee} = e^2/(\epsilon|{\bm r}_1 - {\bm r}_2|)$, where $\epsilon = (\epsilon_{Si} + \epsilon_{SiO_2})/2$, accounting for the effect of the image charge in the oxide.

In a DQD charge fluctuations can cause the height of the potential barrier to fluctuate and introduce random offsets between the bottoms of the two dots.  For single-spin qubits, the exchange coupling $J$ controls interactions between qubits and is pulsed on and off by a gate.  Here we choose a confinement potential $V_x = \frac{\hbar^2}{2m_ta^4} \, \big(\frac{x^2 - X_0^2}{2X_0} \big)^2$, which gives rise to a barrier potential $V_{b1} = (\hbar^2X_0^2/8m_ta^4)$.  In the Heitler-London approximation
\begin{equation}
J = \frac{\frac{3\hbar^2}{4m_ta^2}\, \bigg(1 + \frac{X_0^2}{a^2} \bigg) + \frac{e^2}{\epsilon a}\, [ \mathcal{I} (2X_0) - \, \mathcal{I} (0) ] }{\sinh\big(2X_0^2/a^2\big)},
\end{equation}
where $\mathcal{I} (R) = \int_0^\infty dq \, e^{-\frac{(1 - b^2/a^2)q^2}{2}} J_0 \big(\frac{qR}{a}\big) \, Erfc \big(\frac{bq}{a\sqrt{2}}\big)$, with $J_0$ the zeroth-order Bessel function and $Erfc$ the complementary error function. 

In general the total electrostatic potential that defines a quantum dot contains contributions from all possible sources -- gates, defects, static dopants, etc.  Each source contributes to the barrier as well.  The exchange $J$ is a function of all the contributions and can be written as $J(V_{b1}, V_{b2}, ...)$.  The total fluctuation in $J$ is approximately $\Delta J = \sum_j (\partial J/\partial V_{bj})\, \Delta V_{bj}$, where $j$ labels the different barrier contributions.  The qubit's sensitivity to fluctuations in barrier $j$ is therefore encapsulated in $\partial J/\partial V_{bj}$.  Here we consider only two contributions to the barrier -- from the original confinement potential $V_{b1} = V_x$ and from the trap potential $V_{b2} = V_T$.  While $V_x$ depends on device parameters, $V_T$ is determined by the random distribution of traps. When the trap fluctuates only $V_{b2}$ changes, thus $\Delta J \approx (\partial J/\partial V_{b2})\, \Delta V_{b2}$.  When the potential of the gate electrodes fluctuates to give a nonzero $\Delta V_{b1}$, it causes a fluctuation in the exchange of $\Delta J \approx (\partial J/\partial V_{b1})\, \Delta V_{b1}$.  With $V_{b1}$ and $V_{b2}$ having different geometry and dynamics, and exchange $J$ depends nonlinearly on these variables, generally $\partial J/\partial V_{b2} \neq \partial J/\partial V_{b1}$.

We now estimate the fluctuation $\Delta J$ for a double dot under the influence of RTN, since a fluctuating $J$ leads to a gating error $\int J \, dt$ that results in unwanted entanglement in spin states \cite{HuDS_06}.  For a single DQD we expect 1-4 traps \cite{Wirth_IEEE05RTN}, of which perhaps only one will be close to either dot. We consider a trap at $X_T$ just outside to the left of the left dot, at a depth $d$ in the substrate.  In the Thomas-Fermi approximation, the screened trap potential is
\begin{equation}
V_T ({\bm r}) = \frac{e^2}{4\pi \epsilon_0 \epsilon_r [(x - X_T)^2 + y^2]^{3/2}} \, \bigg(\frac{1 + q_{TF}d}{q_{TF}^2}\bigg),
\end{equation}
where $q_{TF} = 2/a_B$ \cite{Davies}, with $a_B \sim 3$nm the effective Bohr radius in Si.  In the Heitler-London approximation we find $J \rightarrow J + \Delta J$, where $\Delta J = (\Delta v_L + \Delta v_R - 2\Delta w)/\sinh (2X_0^2/a^2)$, with $\Delta v_{L, R} = \tbkt{L,R}{V_T}{L, R}$ and $\Delta w= \frac{\tbkt{L}{V_T}{R}}{\dbkt{L}{R}}$, and we have used $\ket{L,R}$ for the left and right dot wave functions.  We consider a QD with an electron state radius of 8nm. It is expected that a much larger area of radius $\sim$50nm around the dot would be depleted so that any trap in this area will not be charge-active. For a realistic 50 nm interdot separation this gives us $X_T = -75$ nm. With $d$ taken to be negligible and $\epsilon \approx 8$ we find $J \approx 0.4 \mu$eV and $\Delta J/J \approx 7 \times 10^{-5}$. 

Energy-level fluctuations have been observed in GaAs QDs \cite{Jung_APL04}, in Si MOSFETs \cite{Ralls_PRL84, Fleetwood_IEEE02, Wirth_IEEE05RTN, Liu_Private, Zimmerman_JAP08}, and in Si nanowires \cite{Grenoble}.  For a GaAs QD the level fluctuation is found to be $\sim$0.07--0.16$\mu$eV \cite{Jung_APL04}, while for a reasonably quiet Si MOSFET QD it is $\sim$0.45$\mu$eV \cite{Liu_Private}.  However, the length scale of these fluctuations has not been determined at present.  Simultaneous fluctuations in the dot potentials of the same magnitude do not affect the exchange.  In other words, only short wave length fluctuations, which affect the interdot bias or barrier height, affect the value of $J$ and cause dephasing.  Nevertheless, based on the similar values of level fluctuations in GaAs and Si QDs, we conclude that, while Si can have much better one-qubit coherence properties than GaAs, for two-qubit gates it will have similar gate errors due to charge fluctuations.  

\begin{figure}[t]
\includegraphics[width=\columnwidth]{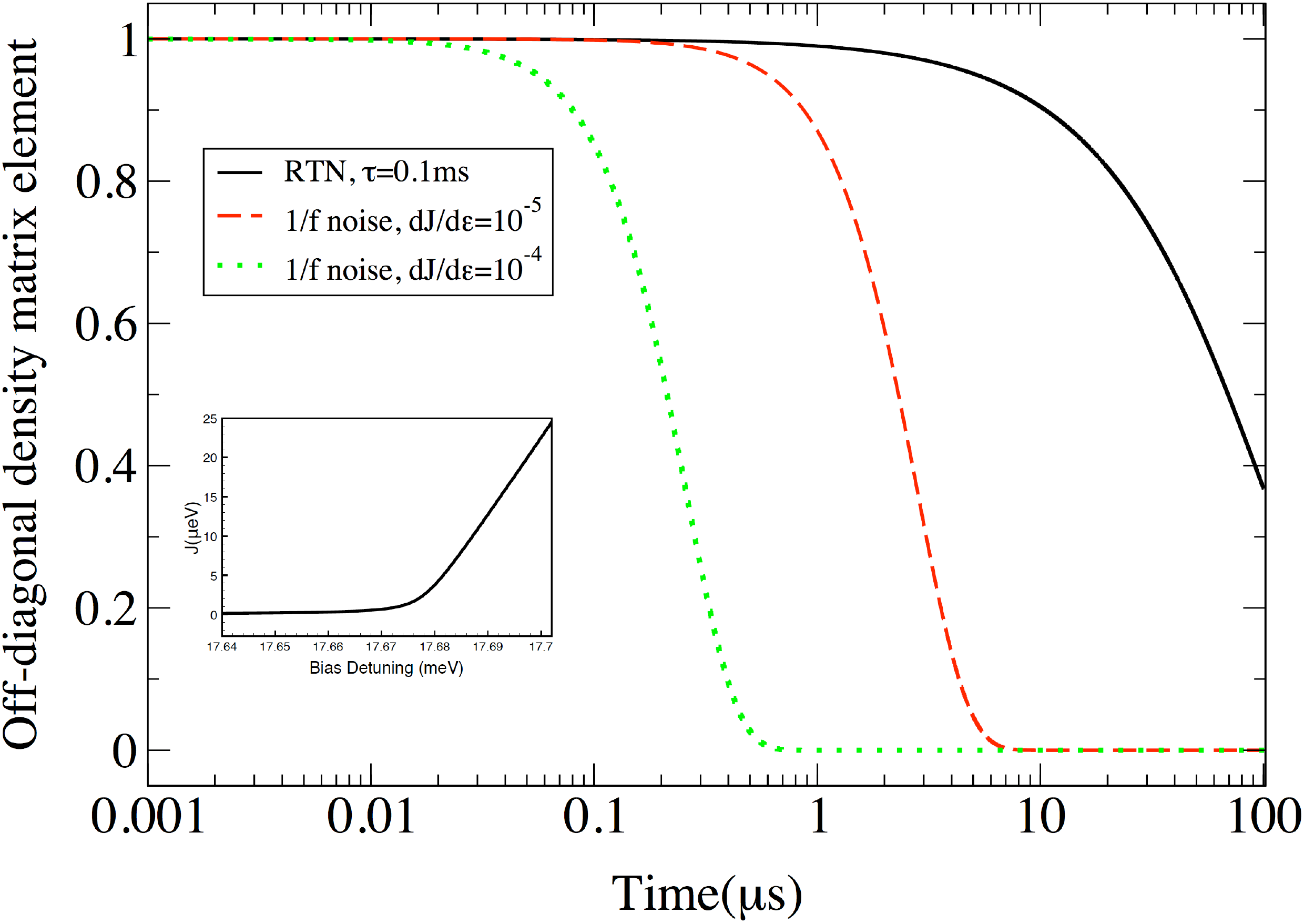}
\caption{Dephasing of a singlet-triplet qubit in a Si/SiO$_2$ DQD with radius $a$ = 8.2nm and $T=$0.1K for RTN and $1/f$ noise. The vertical axis shows the surviving amplitude $\protect\rho_{12}(t)/\protect\rho_{12}(0)$ as a function of time. For RTN we have chosen $\protect\tau = 0.1$ms. For $1/f$ noise the surviving amplitude is shown for different values of $dJ/dV$, with $S_V(\omega) \approx 3 \times 10^{-10}$V$^2$/$\omega$ estimated from Ref.\ \cite{Liu_Private}. The inset shows the exchange coupling $J$ as a function of bias detuning for a DQD with $a$ = 8.2nm and interdot separation 50nm.}
\label{fig:deph}
\end{figure}

A logical qubit can be encoded in the two-electron singlet and unpolarized triplet spin states in a biased regime for a DQD \cite{PettaSci}. In this case, if a finite $J$ is maintained to avoid singlet-triplet mixing by an inhomogeneous magnetic field, charge fluctuations could lead to pure dephasing between the qubit states. To assess this dephasing in Si, we calculate $J$ with a tilted model potential $V_x = (\hbar \omega_0/2a^2) \, \mathrm{Min} [(x-X_0)^2, (x+X_0)^2] - eEx$, where the external electric field $E$ raises the energy of the left dot with respect to the right dot.  Here $J$ is controlled by applying a bias potential energy $\epsilon \sim 2eEX_0$ between the two dots and can be very large, as shown in the inset of Fig.~\ref{fig:deph}. It is given approximately by $J \sim 2\tilde{t}^2/\epsilon$, where $\tilde{t}$ is the tunnel coupling and $\epsilon$ is the energy splitting of the two lowest singlet states.  

In a tilted potential, variations in the exchange are given by $\Delta J = (\partial J/\partial \tilde{t}) \, \Delta \tilde{t} + (dJ/d\epsilon) \, \Delta \epsilon$.  The tunnel coupling is determined mostly by the interdot distance and details of the interdot barrier.  It varies only slightly within the bias regime that we work with, and is treated as an independent variable from $\epsilon$.  Therefore we can write $\Delta J = dJ/d\epsilon \, \Delta \epsilon$. $\Delta \epsilon$ has the same electrical origin as the level fluctuations $\Delta V$ and has contributions from both the electrodes and the charge traps: $\Delta \epsilon = \Delta V_{b1} + \Delta V_{b2}$.  This linear relationship allows us to evaluate the effect of the charge traps based on the gate potentials since $\partial J/\partial V_{b2} = \partial J/\partial V_{b1} = dJ/d\epsilon$.  If we focus only on noise from charge traps, $\Delta J = (dJ/d\epsilon) \Delta V$.  

The singlet and triplet states define a 2D Hilbert space, in which the effective Hamiltonian can be written simply as $H_{eff} = [J + \Delta J(t)] \, \sigma_z$.  In other words, charge fluctuations lead to pure dephasing in the singlet-triplet qubit and a decay in the off-diagonal element $\rho_{12}$ of the system density matrix.
If the double dot qubit is only affected strongly by a few nearby charge traps, the corresponding charge noise can be treated as RTN, which is strongly non-Gaussian.  We find that \cite{SousaHyp} 
\begin{equation}  
\label{deph}
\frac{\rho_{12}(t)}{\rho_{12}(0)} = e^{-t/\tau} \, \big( \cos \eta t  + \frac{1}{\eta \tau} \sin \eta t \big),
\end{equation}
where $\eta = \sqrt{(J + \Delta J)^2/\hbar^2 + (1/\tau^2)}$.  Since $(J + \Delta J)/\hbar \gg 1/\tau$ typically we have $\eta \approx (J + \Delta J)/\hbar$.  Two time scales enter Eq.~(\ref{deph}) -- one associated with precession and one with decay. While the precession time scale is set by $J$ and is fast (shorter than ns), the decay time scale is set by the switching time of the fluctuator. This switching time can be very long and can exceed the spin lifetime in Si.  For example, experiments on a Si/SiO$_2$ nanowire at dilution refrigerator temperatures \cite{Grenoble} have found $\tau$ ranging from 0.1 ms to hundreds of ms depending on the gate voltage. From this perspective working with small samples having few traps is obviously beneficial.

A scalable quantum computer will no doubt be made on large wafers, which are usually subject to $1/f$ noise, typically Gaussian in MOSFETs \cite{Kogan}.  In this case we can write $\rho_{12}(t)/\rho_{12}(0) = e^{-\chi(t)}$, where 
\begin{equation}
\chi (t) = \frac{1}{2\hbar^2}\, \bigg( \frac{d J}{d V} \bigg)^2 \int_{\omega_0}^{\infty}d\omega \, S_V(\omega)\, \bigg(\frac{\sin\omega t/2}{\omega/2}\bigg)^2 \,.
\end{equation}
For $1/f$ noise the low frequency part of the spectrum dominates dephasing, so that $\chi(t) \approx (dJ/dV)^2 (\alpha k_B T/2\hbar^2) \, t^2 \, \ln \omega_0 t$.  The time evolution of the off-diagonal element of the density matrix is given by Fig.~\ref{fig:deph}, where the electron temperature has been set to 100 mK, the typical value in 
a dilution refrigerator.  Figure \ref{fig:deph} shows that dephasing kicks in quite rapidly for the parameter regime we choose, ranging from 0.1 $\mu$s to 10 $\mu$s.  The dephasing time can be enhanced in three ways.  One can go to ever lower temperatures to reduce $S_V(\omega)$, yet this is limited by available technology.  One can also separate the dots further, since larger interdot distances lower $dJ/d\epsilon$.  The best approach may be to apply a magnetic field and find the so-called \textit{sweet spot}, at which $dJ/d\epsilon = 0$ and the double dot is insensitive to charge noise \cite{Stopa_NanoLett08}. Finding such sweet spots may, however, be problematic since quantitatively exact numerical estimates of J(V), taking into account all electrostatic contributions, would be difficult, if not impossible, in a scaled up architecture. 

In summary, we have studied fluctuations in the exchange coupling and dephasing of spin qubits in Si DQDs.  Single-spin qubit systems in Si are expected to have similar gating errors to GaAs.  For singlet-triplet qubit systems there is a significant difference in the two-spin dephasing between submicron devices, in which RTN is expected to be dominant, and large devices, in which $1/f$ noise dominates.  

This work is supported by LPS-NSA and ARO-NSA.

\end{document}